\documentclass[twocolumn,superscriptaddress,showpacs,aps]{revtex4}
\usepackage{graphicx}
\usepackage{epstopdf}
\usepackage{latexsym,amsmath,amssymb}
\usepackage{soul,ulem,balance}
\usepackage[bookmarks=true,pdfpagemode=UseOutlines,bookmarksnumbered=true,colorlinks=true,linkcolor=blue,citecolor=blue,pagecolor=blue,anchorcolor=blue,menucolor=black]
{hyperref}

\renewcommand{\AA}{\overset{\rm{_o}}{\rm{A}}}

\begin{document}

\title{Validity of Stokes-Einstein Relation in Soft Colloids up to the Glass Transition}

\author{Sudipta~Gupta}
\affiliation{JCNS-1 and ICS-1, Forschungszentrum J\"{u}lich, Leo-Brandt-Stra{\ss}e, 52425
J\"{u}lich, Germany}
\affiliation{JCNS-SNS, Oak Ridge National Laboratory, Bethel Valley Road, Oak Ridge, TN 37831, USA}
\author{J{\"o}rg~Stellbrink}
\affiliation{JCNS-1 and ICS-1, Forschungszentrum J\"{u}lich, Leo-Brandt-Stra{\ss}e, 52425 J\"{u}lich, Germany}
\author{Emanuela~Zaccarelli}
\affiliation{CNR-ISC and Dipartimento di Fisica, Universit\'a di Roma La Sapienza Piazzale A. Moro 2, 00185, Rome, Italy}
\author{Christos~N.~Likos}
\affiliation{Faculty of Physics, University of Vienna, Boltzmanngasse 5, 1090 Vienna, Austria}
\author{Manuel~Camargo}
\affiliation{Centro de Investigaciones en Ciencias B\'asicas y Aplicadas, Universidad Antonio Nari{\~n}o, Km 18 via Cali-Jamund\'i, 760030 Santiago de Cali, Colombia}
\author{Peter~Holmqvist}
\affiliation{Division of Physical Chemistry, Lund University, 22100 Lund, Sweden}
\author{J{\"u}rgen~Allgaier}
\affiliation{JCNS-1 and ICS-1, Forschungszentrum J\"{u}lich, Leo-Brandt-Stra{\ss}e, 52425 J\"{u}lich, Germany}
\author{Lutz~Willner}
\affiliation{JCNS-1 and ICS-1, Forschungszentrum J\"{u}lich, Leo-Brandt-Stra{\ss}e, 52425 J\"{u}lich, Germany}
\author{Dieter~Richter}
\affiliation{JCNS-1 and ICS-1, Forschungszentrum J\"{u}lich, Leo-Brandt-Stra{\ss}e, 52425 J\"{u}lich, Germany}

\date{\today, second revised version}


\begin{abstract}
We investigate the dynamics of kinetically frozen block copolymer micelles of different softness across a wide range of particle concentrations, from the fluid to the onset of glassy behavior, through a combination of rheology, dynamic light scattering and pulsed field gradient NMR spectroscopy.  We additionally perform Brownian dynamics simulations based on an ultrasoft coarse-grained potential, which are found to be in quantitative agreement  with experiments, capturing even the very details of dynamic structure factors $S(Q,t)$ on approaching the glass transition. 
We provide evidence that for these systems the Stokes-Einstein relation holds up to the glass transition; given that it is violated for dense suspensions of hard colloids, our findings suggest that its validity is an intriguing signature of ultrasoft interactions.

\end{abstract}
\maketitle


The microscopic origin of vitrification as a system undergoes quenching either by  a sudden decrease in temperature $T$, as in supercooled liquids \cite{Wuttke1994,Angell2000,Lunkenheimer2000}, or by a fast increase of the volume fraction $\phi$, as in colloidal glasses \cite{Pusey1986, Vanmegen1993, Hunter2012}, is still a subject of intense discussion. Colloidal systems have been playing an important role in unveiling the microscopic aspects of the transition due to the capability of tuning specific interactions and in developing experimental and simulation techniques that permit investigations both at the single-particle and the collective levels. Accordingly, it is highly desirable to have at one's disposal a versatile and well-controlled experimental system,  which allows for a systematic tuning of softness at the individual particle level. In addition to fundamental interest, tunable rheological behavior of soft colloids \cite{vlassopoulos:co:2014}  is very important for tailoring material properties with relevance to technological applications in oil and medical industries \cite{Grest1996}. Recently, kinetically-frozen block-copolymer micelles have emerged as easy-to-synthesize and precisely tunable model systems for soft-colloid suspensions \cite{Lund2004, Lodge2004, Laurati2005, Lund2006, Lonetti2011,Gupta2012}. Since the softness and morphology of the micelles can be changed by several factors, which include  the aggregation number, the solvent composition \cite{Lund2004}, the  solvophobic-to-solvophilic block ratio or  the block length, these micelles allow to systematically bridge the physics of linear polymers to that of colloidal hard spheres.
 
One of the fingerprints of a (metastable) liquid approaching the glass transition is the breakdown of the Stokes-Einstein (SE) relation \cite{kob:book}: for glass-forming systems the product of the macroscopic (zero-shear) viscosity $\eta$ and the mesoscopic (long-time) self-diffusion coefficient  $D_s$  is not constant anymore as the system slows down. The origin of this behavior has been attributed to the emergence of dynamic heterogeneities, provided by the presence of distinguishable populations of fast and slow particles \cite{Hodgdon1993, Stillinger1994, Tarjus1995, Debenedetti2001}. The breakdown of the SE relation close to the glass transition is a generic feature shared by atomic  \cite{Bordat} and molecular \cite{ Ngai2000, Ediger2003, Chen2006, Gupta2015} glass formers, polymers \cite{Cicerone} and metallic glasses \cite{chen:metallic}. Interestingly, only a limited number of experimental studies on colloidal systems showing deviations from SE behavior is available, including   hard spheres \cite{Bonn2003} and Laponite suspensions \cite{Jabbari2012}.  Much more abundant in the literature are numerical results, which have allowed to grasp microscopic insights on the nature of the SE breakdown and have shown that both the onset and the strength of the deviations depend on the specific interaction potential. Thus, while atomic and molecular glass-formers inevitably display large deviations upon lowering temperature, colloidal systems, being amenable of a systematic and controlled variation of effective interactions, offer the ideal playground for assessing whether a more extended range of validity of SE exists for specific systems. 

In hard spheres suspensions, systematic SE violations were found both in experiments \cite{Bonn2003} and in simulations \cite{Puertas2007}; moreover, for short-range attractive colloids an enhanced breakdown was observed, a fact that has been rationalized in terms of pronounced dynamical heterogeneities \cite{Puertas2004, Kegel2000, Puertas2007}. In contrast, for  very soft  potentials like the Gaussian core model, no significant violation of SE was observed \cite{Ikeda2011, Coslovich2014}, suggesting that such models are ``mean-field", in the sense that they suppress dynamic heterogeneities due to the presence of long-range interactions. These observations are also very relevant to the recent  discussion of whether there exists special limits, such as high dimensions, where a mean-field description of the glass transition may hold \cite{Biroli2007, Ikeda2011, Schilling2010, Charbonneau2011}. Indeed numerical simulations of hard hyperspheres have also shown a systematic reduction of SE violation with increasing dimensionality \cite{Sengupta2013, Charbonneau2013}. The validity of SE relation up to the onset of glassy arrest may have important consequences also for the experimental community, as it would provide a test-case for the use of advanced microrheological techniques also close to arrest \cite{Squires2010}. Thus, the identification of experimental systems which strictly satisfy SE even in the glassy regime have a broad interest in the scientific community.

In this Letter, we report a comprehensive study of frozen block copolymer micelles featuring {\it tunable softness} \cite{Laurati2005, Lund2004, Lund2006}. Experimental results from rheology measurements, dynamic light scattering (DLS), and pulsed field gradient NMR spectroscopy (PFG-NMR) are compared to those obtained from Brownian dynamics simulations (BD) based on a coarse-grained description of the system. We find excellent agreement between experiments and theoretical description for the dynamic structure factor $S(Q,t)$ over a broad range of volume fractions approaching the glass transition value $\phi_g$, as well as for the density dependence of the transport coefficients, including the self-diffusion coefficient $D_s$, the shear viscosity $\eta$ and the collective relaxation time $\tau$. Most strikingly, we find that the SE relation remains valid in the whole investigated range, up to the glass transition, for two different values of the softness parameter, suggesting that star-like micelles may be viewed as an experimental realization of mean-field glasses, where dynamic heterogeneities are strongly suppressed.

We investigate aqueous solutions of star-like micelles with two representative aggregation numbers covering  the regime from typical ultra-soft ($N_{\rm agg}=120$) to moderately soft ($N_{\rm agg}=500$) colloidal particles. For the first system, the amphiphilic block copolymer poly(ethylene-alt-propylene)-poly(ethylene oxide), PEP-PEO, was employed, while  the second was formed by poly(butyleneoxide)-poly(ethyleneoxide), PBO-PEO \cite{Laurati2005, Gupta2012}.  We covered a broad range of polymer volume fraction  ($0.06\%\le\phi\le4\%$) from  very dilute  suspensions to  well above the glass transition for both systems. The samples' characteristics are summarized in Table \ref{table:sizes}. 

\begin{table}[t]
\centering
\caption{Molecular weight $M_w\,[\rm{kg}/\rm{mol}]$, block ratio $m$:$n$ of PBO to PEO and PEP to PEO repeat units, polydispersity $p$, aggregation number $N_{\rm agg}$, and sizes $R_h\,[\AA]$ (hydrodynamic radius) and $\sigma_{\rm int}\,[\AA]$ (micelle interaction diameter).}
\label{table:sizes}
\begin{small}
\begin{tabular*}{0.485\textwidth}{@{\extracolsep{\fill}}l@{}c@{}c@{}c@{}c@{}c@{}c@{}c@{}c}\hline\hline
  \footnotesize{Diblock}   & $M_{w}$ & $M_{w}$ & $M_{w}$ & \footnotesize{Block}  
          & $p$\footnote[1]{$M_w/M_n$ polydispersity determined by $^{1}$H-NMR and SEC.} 
          & $N_{\rm agg}$
          & $R_h$ & $\sigma_{\rm int}$   \\
  \footnotesize{copolymer}     & \footnotesize{PBO} & \footnotesize{PEP} & \footnotesize{PEO} & \footnotesize{ratio} &         & \\\hline
  \footnotesize{hPBO10-dPEO50} & 10.5 &  $-$ & 52.5 & 1:8  & 1.02 & 500 & 727    & 899\\
    \footnotesize{hPEP1-hPEO20}  & $-$   & 1.1 & 21.9 & 1:33 & 1.04 & 120 & 334    & 380
    \\\hline\hline
 \end{tabular*}
\end{small}
\end{table}

\begin{figure}[!b]
\centering
\includegraphics[width=0.4\textwidth]{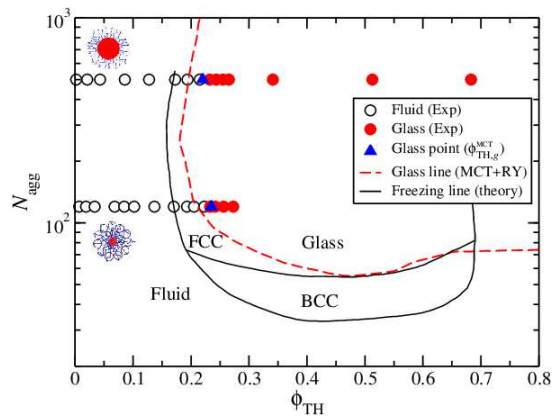}\vspace{-2ex}
\caption{Theoretical phase diagram \cite{Watzlawek1999, Foffi2003} and corresponding experimental path studied in this work. Open and closed symbols indicate investigated fluid and glassy samples, respectively.  Triangles mark the glass transition obtained by extrapolating diffusivity data to zero through a MCT-power-law. The cartoons illustrate the different softness of the micelles at the two studied $N_{\rm agg}$.}
\label{fig:PD}
\end{figure}

To describe the system theoretically,  we employ a coarse-grained model, originally developed for regular star polymers in good solvent conditions \cite{Likos1998, Watzlawek1999}, which was recently shown to be able to describe the structure and phase diagram of star-like micelles \cite{Laurati2005}. In this framework, micelles are coarse-grained as soft particles with an effective, repulsive potential  $V_{\rm eff}(r)$ acting between their centers. The latter features a logarithmic dependence at short distances ($r < \sigma_{\rm int}$) and a Yukawa decay at large ones ($r > \sigma_{\rm int}$). The so-called corona diameter $\sigma_{\rm int}$, roughly equal to the hydrodynamic radius, sets the interaction range while the aggregation number $N_{\rm agg}$, analogous to the star functionality $f$ \cite{Likos1998, Watzlawek1999}, tunes the softness of the repulsion. This model allows to describe the phase behavior of the micellar suspension on the space spanned by $N_{\rm agg}$ and the micellar concentration $\phi_{\rm TH} = (\pi/6)\rho_m\sigma_{\rm int}^3$, where $\rho_m = N_m/V$ is the  density of a system containing  $N_m$ micelles in the volume $V$. While $N_{\rm agg}$ is independently determined by SANS form factor analysis in dilute solution \cite{Lund2004}, $\rho_m$ can be unambiguously determined in experiment in terms of weighted samples. The correspondence between $\rho_m$ and $\phi_{\rm TH}$ is established via the experimental hydrodynamic radius $R_h$, which  is used to determine $\sigma_{\rm int}$, as previously described in equilibrium studies of these systems \cite{Laurati2005}. To study the long-time dynamics of the model, we simulate $N_m=2000$ spherical particles of unit mass $m$, immersed in a cubic box of fixed volume $V_{\rm box}$, interacting through the coarse-grained potential $V_{\rm eff}(r)$. To avoid crystallization at high densities, particle sizes are drawn from a Gaussian distribution of standard deviation 10\% and average diameter $\sigma_{\rm int}$, which sets the unit of length. Simulations are performed at fixed temperature, setting $k_B T=1$ with $k_B$ the Boltzmann constant, for $N_{\rm agg}=120$ and $N_{\rm agg}=500$ at various volume fractions $\phi_{\rm TH}$ controlled by changing $V_{\rm box}$.  To mimic the solvent effectively, we use Brownian dynamics (BD) simulations with the bare, short-time diffusion coefficient fixed to $D_0=0.001$ \cite{noteBD}. The integration time step is chosen as $\Delta t = 0.01$,  where time is measured in units of $\sqrt{m \sigma_{\rm int}^2/(k_BT)}$.

Figure~\ref{fig:PD} displays the phase diagram of the system  in the ($\phi_{\rm TH}$, $N_{\rm agg}$)-plane \cite{Watzlawek1999}, comparing the experimental state points investigated in this work with theoretical results for the fluid-solid \cite{Watzlawek1999} and glass (MCT) \cite{Foffi2003} boundaries. The estimates of the polymer volume fractions at the glass transition, determined by rheology, are $\phi_{g}=(3.476\pm0.03)\%$ and $(2.635\pm0.05)\%$ for $N_{\rm agg}=120$ and $500$ respectively, corresponding to micellar packing fractions $\phi_{{\rm TH},g}\simeq 0.236\pm0.006$ and $0.230\pm0.010$, in good agreement with MCT predictions \cite{Foffi2003}. 

\begin{figure}[!b]
\centering
\includegraphics[width=0.38\textwidth]{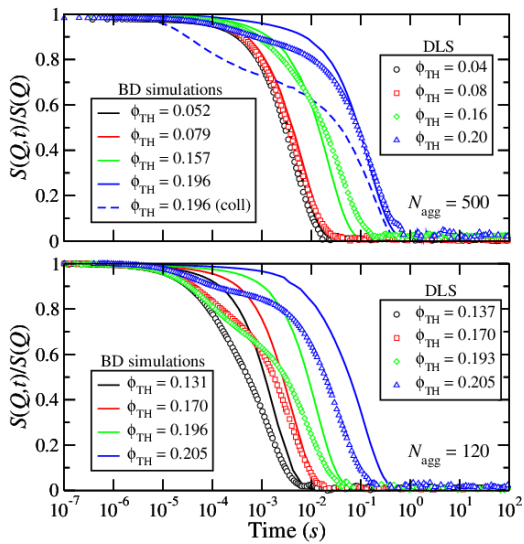}\vspace{-2ex}
\caption{Time dependence of normalized dynamic structure factors  from DLS (symbols) compared with self correlators calculated from BD simulation (solid lines). Top panel: Results for $Q\sigma_{\rm int} \simeq 1.68$ and $N_{\rm agg}=500$. As a reference, also the corresponding collective correlator (dashed line) calculated from BD at the same $Q\sigma_{\rm int}$ is reported for the largest $\phi_{\rm TH}$. Bottom panel: Same as before for $Q\sigma_{\rm int} \simeq 1.07$ and $N_{\rm agg}=120$. Simulation data were multiplied by a constant factor of $11~\mu$s and $2~\mu$s for $N_{\rm agg}=500$ and $120$, respectively, to match the experimental microscopic relaxation.}
\label{fig:Sqt_phi}
\end{figure}

Let $S(Q,t) = N^{-1}\langle \sum_{ij} \exp[-{\rm i}{\bf Q}\cdot({\bf r}_i(t)-{\bf r}_j(0))]\rangle$ be the dynamical structure factor of a system of $N$ particles, ${\bf r}_k(t)$ denoting the position of particle $k$ at time $t$, and $S(Q)$ the equal-times, static structure factor. A detailed comparison between the ratios $S(Q,t)/S(Q)$ measured by DLS and those calculated from BD simulations is reported in Fig.~\ref{fig:Sqt_phi} for different volume fractions at fixed  wavevectors  $Q\sigma_{\rm int}\simeq 1.68$ and $1.07$ for $N_{\rm agg}=500$  and 120, respectively. To superimpose simulation data onto the experimental timescale, the microscopic dynamics is adjusted through an arbitrary shift of the time-axis, which depends on $N_{\rm agg}$ but is the same for all volume fractions. As $\phi_{\rm TH}$ increases, the typical pattern of glass-forming systems emerges in $S(Q,t)$: a two-step relaxation develops with a growing plateau at intermediate times followed by a long-time final relaxation, which grows on approaching the glass transition. The experimentally measured correlators are found to be somehow intermediate (e.g., in the value of the plateau) between the self and collective numerical ones (see Supplementary Information \cite{SM2015}). However, the long-time relaxation is well described by both the self and collective numerical data, capturing quantitavely the growth of the relaxation time with increasing $\phi_{\rm TH}$ and the shape (i.e., the stretching exponent) of the correlators. We stress that apart from the adjustment  of the microscopic time, there is no fit parameter in the comparison, yielding an almost quantitative description  within experimental error for both studied values of $N_{\rm agg}$.

\begin{figure}[!t]
\centering
\includegraphics[width=0.475\textwidth]{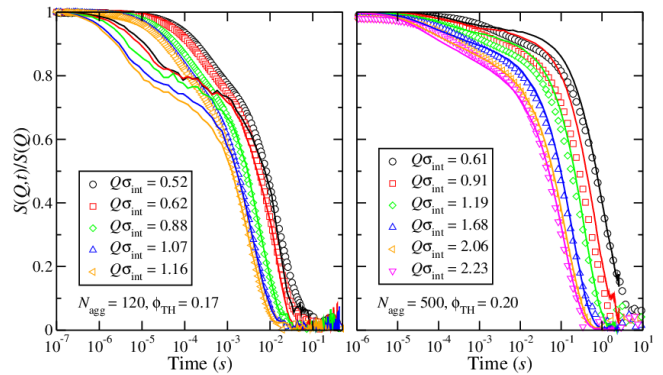}\vspace{-2ex}
\caption{Wavevector dependence of the normalized dynamic structure factors from DLS (symbols) and collective density correlators from BD simulation (solid lines) for $\phi_{\rm TH}=0.17$ and $N_{\rm agg}=120$ (left panel), and for $\phi_{\rm TH}=0.2$ and $N_{\rm agg}=500$ (right panel).}
\label{fig:Sqt_q}
\end{figure}

Additional evidence for the accuracy of the coarse-grained interaction to describe the equilibrium dynamics is provided by the comparison of the $S(Q,t)/S(Q)$  from experiments and simulations as a function of the scattering vector $Q$, which is shown in Fig.~\ref{fig:Sqt_q} at constant volume fraction. Clearly, the $Q$-dependence of the slowest relaxation process in terms of characteristic time and shape (stretching exponent), as well as the $Q$-dependence of the intermediate non-ergodicity plateau are well described by BD simulations for both values of $N_{\rm agg}$, covering a length scale variation between $0.5 \lesssim Q\sigma_{\rm int} \lesssim 2.2$; as a reference, close to the glass transition the nearest-neighbor peak of $S(Q)$ is found at $Q\sigma_{\rm int}  \simeq 5$.
 
\begin{figure}[t!]
\centering
\vspace{-0.2cm}
\includegraphics[width=0.48\textwidth]{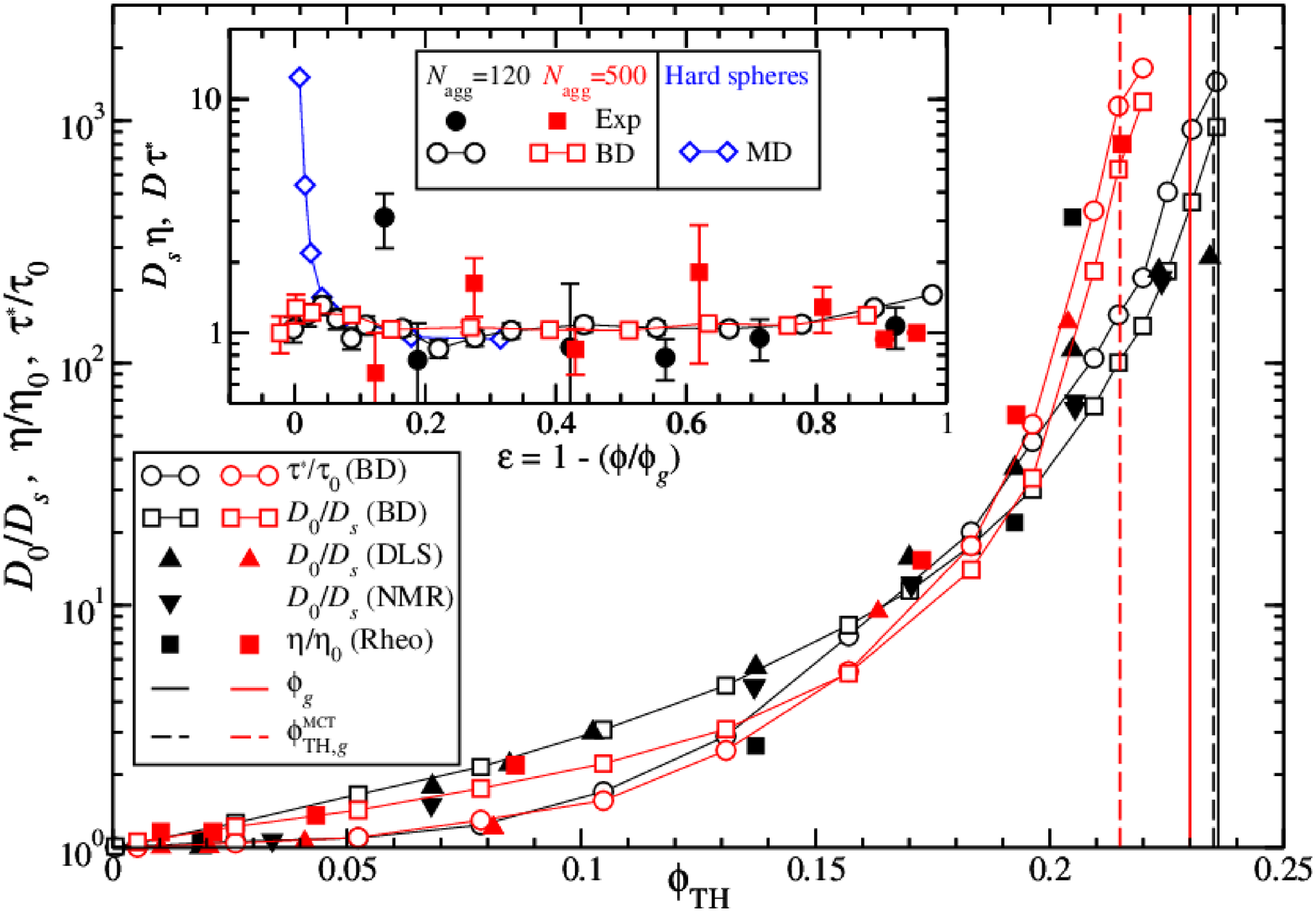}\vspace{-2ex}
\caption{Inverse of reduced self-diffusion coefficient $(D_0/D_s)$ from DLS, PFG-NMR and BD simulations, reduced zero-shear viscosity $(\eta/\eta_0)$ from rheology, and   reduced relaxation time  $(\tau^*/\tau_0)$ calculated at $Q\sigma_{\rm int}=5.0$  from BD simulations, are shown together as a function of volume fraction $\phi_{\rm TH}$ for $N_{\rm agg}=500$ (black) and $N_{\rm agg}=120$ (red).  The experimental glass transition for both $N_{\rm agg}$ are indicated by the vertical solid lines. The corresponding MCT glass lines, obtained by power law fits of the numerical data, are indicated by vertical dashed lines. Inset:  $D_s\eta$ from experiments and $D_s\tau^*$ from simulations as function of the distance to the glass transition $\epsilon$. Collective relaxation times $\tau^*$ were calculated by fitting the correlators at the peak of $S(Q)$ to stretched exponentials. Hard-sphere data for $D_s\eta$  were obtained by MD simulations as explained in Ref.~\cite{Puertas2007}.}
\label{fig:gse}
\end{figure}

So far we have shown that the coarse-grained potential \cite{Likos1998} can describe both the structure \cite{Laurati2005, Lonetti2011} and the dynamics, in a semi-quantitative way, of frozen block copolymer micelles. Building on this, we now provide evidence of the validity of SE relation in these systems by combining direct experimental measurements and information that can be extracted from the theoretical description of the dynamical correlators. To this end, we have investigated the concentration dependence of the long-time, self-diffusion coefficient $D_s(\phi)$ by means of both DLS and PFG-NMR, and also the zero-shear viscosity $\eta(\phi)$ through rheological measurements. In Fig.~\ref{fig:gse}  the reduced viscosity $\eta(\phi)/\eta_0$ and the inverse of the reduced self-diffusion coefficient $D_0/D_s(\phi)$ are reported as a function of $\phi_{\rm TH}$, approaching the glass transition for both experimental systems with $N_{\rm agg}=120$ and $500$. Here $\eta_0$ represents the solvent viscosity, whereas $D_0$ is the diffusion coefficient of the micelles at infinite dilution. For comparison, in Fig.~\ref{fig:gse}  we also report results for the long-time diffusion coefficient $D_s$ and the collective structural relaxation time $\tau$ extracted from BD simulations. Exploiting the fact that the viscosity is proportional to the relaxation time, we present data for the latter at fixed wavevector $Q\sigma_{\rm int}=5$ and normalized by the corresponding time at infinite dilution, $\tau_0$.  Excellent overlap (within experimental error) between measured data and results from BD simulation based on the coarse-grained model is found  without the introduction of any adjustable parameter. Furthermore, we establish that up to the largest 
volume fractions in Fig.~\ref{fig:gse}, the self-diffusion coefficient, the viscosity and the relaxation time for each of the two studied systems follow each other very closely, approaching the glass transition in a similar way. 

These findings suggest that SE holds for such soft block copolymer micelles with star-like interactions. To highlight this, in the inset of Fig.~\ref{fig:gse} we plot the SE products $D_s\eta$ and $D_s\tau^*$, where $\tau^*$ is the collective relaxation time calculated at the peak of $S(Q)$, which is found to give the dominant contribution to viscosity \cite{Puertas2007}. Results are shown there from available experimental and numerical data points, as a function of the distance to the glass  transition $\epsilon$; the latter being defined as $\epsilon=1-\phi/\phi_g$ for experiments and as $1-\phi_{\rm TH}/\phi_{{\rm TH},g}^{\rm MCT}$ for simulations. Here, $\phi_{{\rm TH},g}^{\rm MCT} = 0.235\pm 0.010$ and $0.215\pm 0.010$ are the extrapolated MCT glass transitions using a power-law fit for $N_{\rm agg}=120$ and $500$, respectively \cite{NOTEMCT}.  While data suffer from statistical noise, it is clear from both sets of data that no systematic deviations from SE up to the largest studied volume fractions are observed. These results cover a range which even exceeds the MCT transition ($\epsilon < 0$) for the BD data, thus SE validity is preserved even at very large degree of supercooling where deviations should normally be large.  It is interesting to note that this behavior is  consistent with previous observations that star-polymer glasses, differently from other colloidal glasses, undergo a cessation of the aging process after a certain time $(\sim10^4~\rm{s})$ \cite{Christopoulou2009} and in fact achieve equilibrium, which could provide a physical explanation of the suppression of dynamic heterogeneities in these systems. Our findings for an ultrasoft coarse-grained model, which quantitatively reproduces the structural and dynamical behavior of frozen block copolymer micelles as a realization of star-like soft colloids, are in agreement with those of simulations of the Gaussian core model, another effective potential often used to describe the behavior of soft particles \cite{Ikeda2011}, but are in stark contrast with both experimental and numerical results on hard-sphere colloids \cite{Bonn2003, Puertas2007}. Thus, it appears that softness is at least a necessary requirement  in colloidal systems for the suppression of dynamical heterogeneities and the persistence of the validity of SE relation close to the glass transition.

To summarize, we have examined the dynamics of soft colloids in the vicinity of the glass transition by a combination of experimental techniques for different values of particle softness. Through BD simulations of a coarse-grained, ultrasoft model \cite{Likos1998}, we have been able to accurately describe the measured dynamic structure factors $S(Q,t)$, both as a function of the micellar volume fraction and as a function of the wavevector, for different values of micellar aggregation number. We also observed that the increase of the macroscopic viscosity and of the inverse mesoscopic, self-diffusion coefficient is quantitatively captured by the theoretical data.  The coherent increase of both quantities indicates no violation of the SE relation up to the glass volume fraction. Our findings impressively confirm the microscopic origin of colloidal ``softness" and its effects on the validity of the Stokes-Einstein relation for degrees of metastability for which it normally breaks down in the case of hard colloidal and molecular systems. In this way, they open up new realms for understanding and tailoring complex fluids not only with respect to structure and phase behavior \cite{Laurati2005} but also for colloidal dynamics \cite{Gupta2015b}.

S.G.~and J.S.~acknowledge support from the International Helmholtz Research School (IHRS) Bio-Soft and the DFG within the SFB-TR6, E.Z.~from the MIUR-FIRB ANISOFT (RBFR125H0M), and M.C.~from FPIT (Banco de la Rep\'ublica,  Convenio 201312). E.Z.~and~C.N.L.~acknowledge financial support from  ETN-COLLDENSE (H2020-MCSA-ITN-2014, Grant No.\ 642774).


\begin{thebibliography}{99}
\bibitem{NOTEMCT} These values are moderately larger than those obtained solving MCT long-time equations within the Rogers-Young approximation, corresponding to $\phi^{\rm MCT,RY}_{{\rm TH},g}=0.207$ and $0.190$ for $N_{\rm agg}=120$ and $500$, respectively.
\end{thebibliography}
\end{document}